\documentclass[smallextended]{svjour3}  

\smartqed 
\usepackage{graphicx}
\usepackage{xcolor}
\usepackage{amsmath}
\usepackage{amsfonts}
\usepackage{amssymb}
\usepackage[labelfont=bf,labelsep=space]{caption}
\usepackage{subcaption}
\usepackage{bbm}
\usepackage{algorithm}
\usepackage{algpseudocode}
\usepackage[english]{babel}
\usepackage{cite}

\begin{document}

\title{Enhancing Quantum Annealing Performance for the Molecular Similarity Problem}
\titlerunning{Enhancing Quantum Annealing Performance} 
\author{Maritza Hernandez         \and
            Maliheh Aramon 
}

\institute{	
	$^{*}$ Both authors have contributed equally to this work.\\   \\       
	Maritza Hernandez \at
              1QB Information Technologies (1QBit), Vancouver, British Columbia, Canada  V6B 4W4\\
              \email{maritza.hernandez@1qbit.com}           
           \and
           Maliheh Aramon \at
           1QBit, Vancouver, British Columbia, Canada  V6B 4W4\\
           \email{maliheh.aramon@1qbit.com}  
}


\maketitle
\begin{abstract}
Quantum annealing is a promising technique which leverages quantum mechanics to solve hard optimization problems. Considerable progress has been made in the development of a physical quantum annealer, motivating the study of methods to enhance the efficiency of such a solver. In this work, we present a quantum annealing approach to measure similarity among molecular structures. Implementing real-world problems on a quantum annealer is challenging due to hardware limitations such as sparse connectivity, intrinsic control error, and limited precision. In order to overcome the limited connectivity, a problem must be reformulated using minor-embedding techniques. Using a real data set, we investigate the performance of a quantum annealer in solving the molecular similarity problem. We provide experimental evidence that common practices for embedding can be replaced by new alternatives which mitigate some of the hardware limitations and enhance its performance. Common practices for embedding include minimizing either the number of qubits or the chain length, and determining the strength of ferromagnetic couplers empirically. We show that current criteria for selecting an embedding do not improve the hardware's performance for the molecular similarity problem. Furthermore, we use a theoretical approach to determine the strength of ferromagnetic couplers. Such an approach removes the computational burden of the current empirical approaches, and also results in hardware solutions that can benefit from simple local classical improvement. Although our results are limited to the problems considered here, they can be generalized to guide future benchmarking studies.

\end{abstract}

 \begin{acknowledgements} 
 This research was supported by 1QBit. The authors would like to thank Clemens Adolphs,  Hamed Karimi,  Anna Levit, Dominic Marchand, and Arman Zaribafiyan for useful discussions, Robyn Foerster for valuable support, and Marko Bucyk for editorial help. We thank Helmut Katzgraber for reviewing the manuscript. We also acknowledge the support of the Universities Space Research Association (USRA) Quantum Artificial Intelligence Laboratory Research Opportunity program.
\end{acknowledgements}

\section{Introduction}
\label{sec:intro}
Quantum annealing (QA), the quantum counterpart of simulated annealing, is an approach for harnessing quantum mechanical effects in searching the energy landscapes of classical NP-hard optimization problems \cite{Santoro06-QA,Farhi01-QA}. The development of a quantum annealer by \emph{D-Wave Systems} \cite{Johnson11-DW} has initiated a great deal of theoretical and experimental research into the usefulness of the QA approach and its potential supremacy over classical algorithms~\cite{Denchev15-Google,Hen15-Probing,Venturelli15-SKM,Vinci15-Error,Katzgrabe15-Ugly,Zhu16-Yield,Perdomo16-Calibration}. 

In recent years, the research community has focused mainly on searching for a useful area of application where QA demonstrates a scaling advantage over classical algorithms. Although there is evidence that the D-Wave quantum annealer exhibits quantum means of energy landscape exploration such as tunnelling \cite{Boixo16-Tunneling} and entanglement \cite{Lanting14-Entanglement}, the efforts to identify an application for which the device is able to outperform classical optimization have not yet been conclusive. The highlight of the search for potential quantum supremacy is the recent study by Google \cite{Denchev15-Google} on an artificially crafted weak--strong cluster problem showing that quantum approaches using either the D-Wave 2X (DW2X) quantum annealer or quantum Monte Carlo simulation scale significantly better than simulated annealing. 

The primary difficulty in demonstrating a quantum scaling advantage on useful optimization problems can be attributed to the architecture of the quantum annealer. The quantum annealer is designed to find the ground state of Ising Hamiltonians with pair-wise interactions on a fixed sparse graph called ``Chimera''. Although the majority of optimization problems across various disciplines can be translated into Ising problems, their formulations usually have connectivity different from that specified by the Chimera's structure. Minor embedding (ME) is a technique for mapping such \emph{non-native} problems to the hardware, where several physical qubits encode one logical qubit \cite{Cai14-ME,Choi11-ME}.

Due to the ME overhead, embedded problems are suboptimal for determining a scaling 
advantage \cite{Mandra2016}. However, the analysis of the quantum annealer's performance on these problems is critical for 
the design of future hardware architectures and the setting of various programming 
parameters. Examples of studies on parametrized families of hard embedded problems include the Sherrington--Kirkpatrick 
model with random $\pm1$ couplings, which is directly related to the graph partitioning problem \cite{Venturelli15-SKM}, operational 
navigation and scheduling problems \cite{Rieffel15-Planning}, the job-shop scheduling problem \cite{Venturelli15-JSP}, the multi-period portfolio 
optimization problem \cite{Rosenberg16-Portfolio}, and the graph isomorphism problem \cite{Zick15_GI}. In this paper, we report on the 
performance of the DW2X and discuss efficient programming guidelines for the problem of measuring similarity among 
small molecules that are modelled as labelled graphs. To determine the similarity between two molecular graphs while accounting for 
noise, a relaxation of the maximum weighted independent set problem, known as maximum weighted co-$k$-plex problem, is formulated 
such that it is consistent with the hardware's architecture. Previous studies of non-native 
problems were conducted using randomly generated instances. To the best of our knowledge, our work is the first to 
examine the performance of a quantum annealer on real instances of problems in the context of molecular similarity. 

Encoding and decoding are challenges specific to solving non-native problems on a quantum annealer. 
Encoding includes two problems: topological embedding and parameter setting. In the former, 
the mapping between each logical qubit and a set of connected physical qubits is determined. In the latter, 
the strength of internal couplings (among physical qubits corresponding to the same logical qubit) is set, and 
the logical local fields and coupling values are distributed among the physical qubits and couplers. It is well known that both topological 
embedding and parameter setting problems have a significant impact on the efficiency of a quantum annealer \cite{Rieffel15-Planning}. 
Decoding refers to the process of inferring the solution of each logical qubit from the retrieved 
solutions of the corresponding physical qubits. An important consequence of the limited available precision \cite{Choi08-Parameter} 
and the existence of errors \cite{Venturelli15-SKM} of the quantum annealer is that the physical qubits representing a logical qubit 
might be assign to different values. Classical post-processing techniques are often used to assign the right value to 
the logical qubit. In this paper, we review current approaches for encoding and decoding 
non-native problems and provide alternatives to the current suboptimal practices, which require substantial computational time.

The rest of the paper is organized as follows. In Section~\ref{sec:molecular_similarity}, we define the molecular 
similarity problem. In Section~\ref{sec:quantum_annealing}, a background on QA is presented and a novel 
formulation of the molecular similarity problem amenable to QA is developed. Different challenges of solving 
an embedded problem on the hardware architecture are then discussed. Detailed experimental 
results are presented in Section~\ref{sec:experimental_results}. We conclude and discuss 
future work in Section~\ref{sec:conclusion}. Supplementary information is presented 
in the Appendix.

\section{Molecular Similarity}
\label{sec:molecular_similarity}
The measurement of structural similarity among molecules plays an important role in 
chemical and pharmaceutical research in areas such as drug discovery. 
The similarity measures proposed in the literature can be categorized into two classes. 
The first class uses a vector-based representation called a fingerprint in which the molecules 
are compared using distance metrics such as Euclidean distance \cite{Baum07}. 
Although fingerprints are simple and computationally efficient, they cannot provide accurate 
information on the common substructures of molecules. Thus, in this paper, we use 
the other class, which was developed based on graph theoretical concepts. 

The second class of similarity measures uses the intuitive graph representation of the labelled molecular atom--bond 
structure. Formally, a labelled graph of a molecule can be written 
as $G=(V,E,\mathcal{L}_{V},\mathcal{L}_{E})$, where $V$ is the set of vertices, $E \subset V \times V$ is the set of edges, 
$\mathcal{L}_{V}$ is the set of labels assigned to each vertex, and $\mathcal{L}_{E}$ is the set of edge labels. 
Graph representations of molecules are often compared based on the property of isomorphism. 
Two graphs $G_1$ and $G_2$ are called isomorphic if there is a bijection between their vertex sets such that there exists 
a mapping between the adjacent pairs of vertices of $G_1$ and $G_2$. A more practical variation of isomorphism is the maximum weighted common subgraph (MWCS), which identifies the largest weighted subgraph of $G_1$ that is isomorphic to a subgraph of $G_2$. There is a correspondence between the MWCS problem and another well-known problem---the maximum weighted independent set (MWIS) of a third graph, which can be induced from the graphs being compared \cite{Khazm07,Balasundaram13}. The vertices and edges of the third graph, called a \emph{conflict graph}, represent possible mappings and the conflicts between them, respectively. The goal of the MWIS problem is to find the largest weighted set of vertices such that there is no edge between all selected pairs, forming the largest conflict-free mapping. 

\begin{figure}[h!]
\centering
\includegraphics[width=0.8\linewidth]{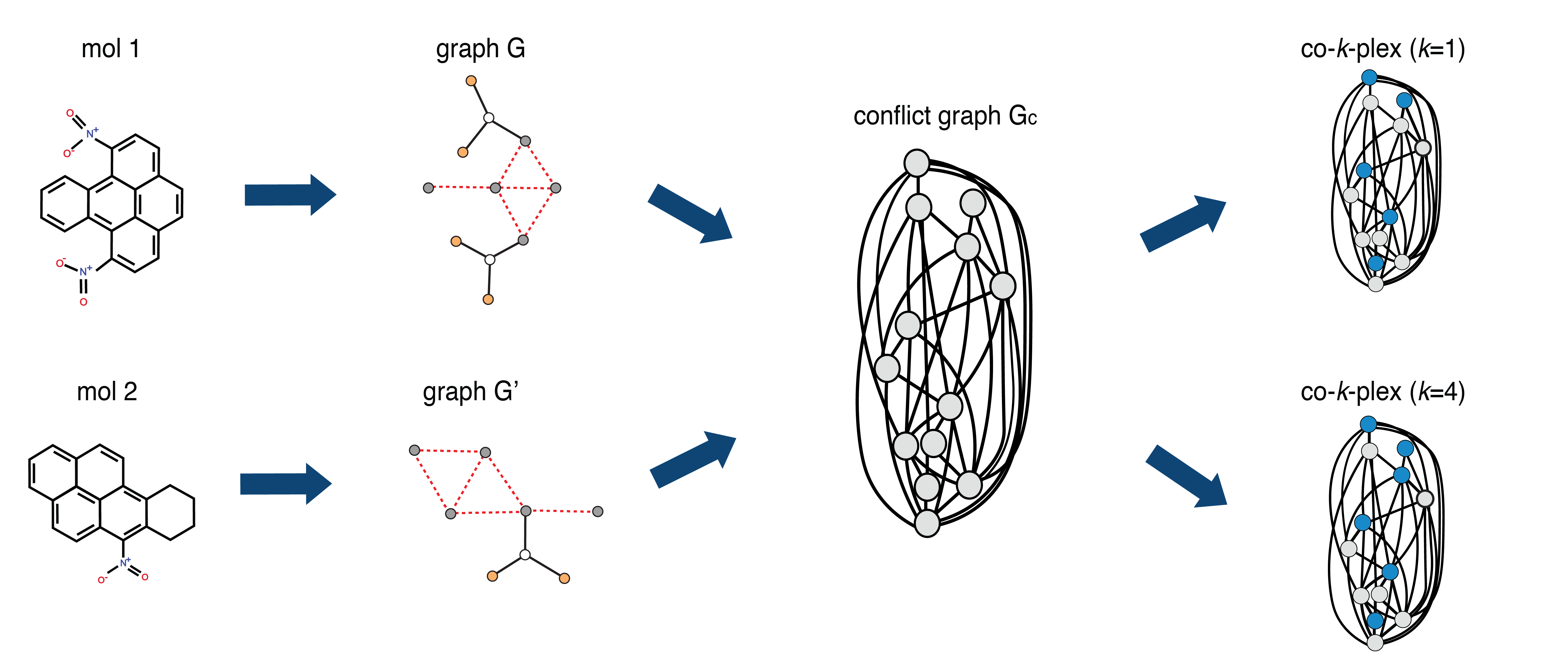}
\caption{Illustration of the graph-based molecular similarity problem. Two molecules are represented as graphs and a conflict graph $G_{\mbox{c}}$ is built. The maximum weighted co-$k$-plex solution is shown for two cases, $k=1$ and $k=4$. }
\label{fig:cokplex_example}
\end{figure}

Since molecular data are subject to regular errors, representing the MWCS problem as the MWIS problem makes it easier to relax the definition of similarity to account for the effect of noise in the data by considering as similar substructures 
with conflicts up to a certain threshold. There are different relaxations of the similarity requirement in 
the literature \cite{k_plex_social}. One of the relaxations is known as the maximum 
weighted co-$k$-plex problem, in which the goal is to find the largest weighted set of vertices in the graph such that each 
vertex has at most $k-1$ edges connecting it to the other vertices. It is clear that the maximum weighted 
co-$1$-plex problem is the MWIS problem \cite{Balasundaram13}. The majority of the similarity methods discussed above, 
including the MWIS and maximum weighted co-$k$-plex problems, are in general NP-hard, having exponentially increasing 
computational complexity due to the combinatorial nature of the graphs involved \cite{Garey79,Downey95}. 
An illustration of the molecular similarity problem and its co-$k$-plex formulation is shown in Figure \ref{fig:cokplex_example}.

The details of reducing molecules to graphs and building the corresponding conflict graph are discussed in our previous work \cite{Hernandez16-GS}. In the next section, we present an overview on 
QA and discuss how the problem of measuring molecular similarity can be solved by a quantum 
annealer.

\section{A Quantum Annealing Approach to the Molecular Similarity Problem}
\label{sec:quantum_annealing}

Quantum annealing is a heuristic technique that was introduced to solve hard optimization problems by exploiting quantum mechanical effects such as quantum tunnelling. The solution to a combinatorial optimization problem is encoded into the ground state of a classical Ising Hamiltonian
\begin{equation}
H_{ \text{Ising}} = \sum_{i \in V} h_{i} s_{i} + \sum_{(i,j) \in E}J_{ij} s_{i} s_{j}, \quad s_{i} \in \pm 1,
\label{eq:ising}
\end{equation}
where the local fields and couplers are represented by $ h_{i} $ and $ J_{i j} $, respectively. The variables $s_{i}$ denote classical Ising spin variables and the sums run over the weighted vertices $V$ and edges $E$ of a graph $G = (V, E)$. Specifically, the task is to find the spin configuration $ \{ s_{i}  \} $ which minimizes $H_{ \text{Ising}}$. A quantum annealing solver attempts to minimize $H_{ \text{Ising}}$ by implementing a time-dependent Ising Hamiltonian
\begin{equation}
H(\tau) = - A(\tau) \sum_{i} \sigma_{i}^{x}+ B(\tau) H_{ \text{Ising}}, \quad \tau \in [0,1],
\end{equation}
where $ \sigma_{i}^{x}$ is the Pauli spin operator on qubit $i$ and  $\tau = t / t_{a}$, with $t_{a}$ referred to as the annealing time. The functions $A(\tau)$ and $B(\tau)$ specify the annealing schedules where typically $A(\tau)$ and $B(\tau)$ are monotonically decreasing and increasing functions, respectively. 

The D-Wave devices accept problems in terms of the Hamiltonian described in Equation~\ref{eq:ising}. However, in conventional optimization formulations, binary variables $z_{i}$ commonly take values from $\{ 0,1 \}$ instead of $\{ -1,1 \}$. This mathematical form is usually referred to as a quadratic unconstrained binary optimization (QUBO) problem. Fortunately, the QUBO and Ising formulations can be related by taking $s_{i} = 1 - 2 z_{i}$. Thus, to solve any optimization problem on a quantum annealer, it is sufficient to reformulate it as an instance of QUBO. In the sections to follow, we describe how to formulate the maximum weighted co-$k$-plex problem in QUBO form, how to map a QUBO problem to the quantum hardware, and how to retrieve the logical answers from the physical solutions provided by the quantum annealer.

\subsection{QUBO Representation}
\label{subsec:qubo_representation}
A co-$k$-plex of a graph $G$ is a subgraph of $G$ in which each 
vertex has a degree of at most $k-1$. Formally, let $G_{\mbox{c}}=(V_{\mbox{c}}, E_{\mbox{c}})$ be the conflict graph 
of two molecular graphs $G_1$ and $G_2$, where $V_{\mbox{c}} \subset V_1 \times V_2$, and $E_{\mbox{c}}$ represents both 
the bijection and the user-defined requirements. The latter allows the enforcement of customized 
structure restrictions, for example, adding an edge to $E_{\mbox{c}}$ if there is a mismatch between the 
edge labels of two molecular graphs. The maximum weighted co-$k$-plex of the 
conflict graph $G_{\mbox{c}}$ corresponds to the MWCS of $G_1$ and $G_2$, 
where each possible pairing can violate at most $k-1$ constraints. Before presenting the QUBO 
formulation, let us first define a star graph.
\\
\begin{definition}
\label{stardef}
A graph $S^k=(V, E)$ is a star graph of size $k$ if it is a tree with $k+1$ vertices and one vertex of degree $k$. 
\end{definition}

Based on the co-$k$-plex formulation, we do not penalize the conflict edges. Each vertex in the conflict 
graph can have up to $k-1$ edges. Therefore, we only penalize in situations where a subset of the vertices 
induces a subgraph in which there is one vertex with degree greater than $k-1$. In other words, 
we penalize all subsets of vertices whose induced subgraph forms a star graph of size $k$.

Further, let us define the binary parameter $\mathcal{A}_{v_1, \ldots, v_{k+1}}$ as
\begin{align}
\mathcal{A}_{v_1, \ldots, v_{k+1}} = \begin{cases}
1   & \text{if~} \{v_1, \ldots, v_{k+1} \} \text{~induces~} S^{k}, \\
0   & \text{otherwise}, \notag
\end{cases}
\end{align}
where $v_1, \ldots, v_{k+1}$ are the vertices of the conflict graph $G_{\mbox{c}}$.

The maximum weighted co-$k$-plex problem is formulated as
 
\begin{align}
\label{eq:co-k-plex_QUBO}
\mbox{max~~} \Bigg[ & \sum_{\substack{v_i \in V_{\mbox{c}}}} w_{v_i} x_{v_i} - \Big(\sum_{\substack{(v_1,\ldots,v_{k+1})}} a_{v_1,\ldots,v_{k+1}} 
                                           \mathcal{A}_{v_1,\ldots,v_{k+1}} \prod\limits_{i=1}^{k+1}x_{v_i} \Big) \Bigg] ,
\end{align}
where $x_{v_i}$ is a binary variable equal to 1 if the vertex $v_i$ is included in the maximum independent set 
or 0 otherwise, $w_{v_i}$ is the weight of vertex $v_i$, and $a_{v_1, \ldots, v_{k+1}} > \mbox{min} \{w_{v_1},\ldots,w_{v_{k+1}}\}$. 
The tunable parameter $k$ should be determined by the user. 

The objective function of Formulation \eqref{eq:co-k-plex_QUBO} is a 
higher-order polynomial. There are several algorithms in the literature that map higher-order polynomials 
to quadratic polynomials \cite[and references therein]{hobo2qubo}.

\subsection{Encoding a QUBO Problem on the Quantum Processor}
\label{subsec:optimal_mapping}
The hardware architecture of the D-Wave devices consists of a fixed sparse graph called a Chimera graph, in which each vertex is a physical flux qubit and each edge is a physical coupler between two qubits. Each generation of devices has shown significant progress, including an increase in the number of qubits and reduction of noise. However, implementing a practical problem remains a challenge, mainly because of the physical constraints of the hardware. One of the fundamental limitations of these devices is their sparse connectivity.

\begin{figure}[h!]
\centering
\includegraphics[width=0.7\linewidth]{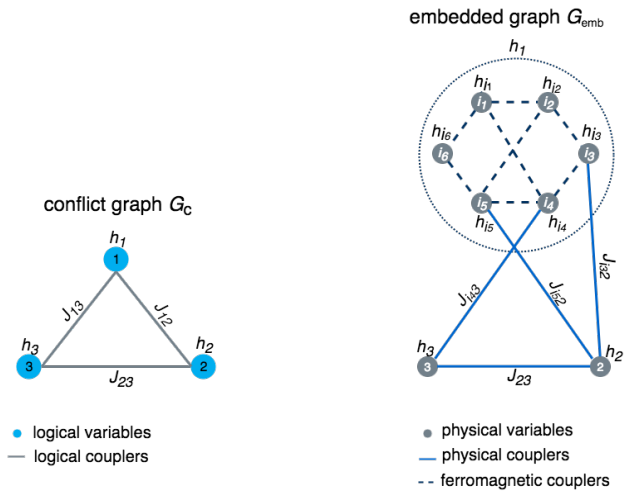}
\caption{An example of embedding a conflict graph $G_{\mbox{c}}$ into the hardware graph. The logical 
variable $1$ is replaced by a connected subgraph with physical vertices $\{i_1, \ldots, i_6\}$ which are 
ferromagnetically coupled.}
\label{fig:embedding_example}
\end{figure}

Typically, a real-world problem formulation might require a different connectivity from that defined by the hardware graph. This is true of the molecular similarity problem, which involves finding the maximum weighted co-$k$-plex in a conflict graph $G_{\mbox{c}}$. Therefore, we need to encode the problem into the device's architecture. There are two aspects to this encoding process: the embedding problem and the parameter-setting problem. The embedding problem is commonly addressed by ME techniques where each vertex of $G_{\mbox{c}}$ is replaced by a connected subgraph of the hardware graph, denoted by $S_{i} = (V_{S_i},E_{S_i})$. Each connected subgraph is ferromagnetically coupled and is referred to as a chain in the literature, though it does not necessarily have a linear, acyclic structure. An example of ME is illustrated in Figure \ref{fig:embedding_example}, and more details about ME can be found elsewhere \cite{Cai14-ME,Choi11-ME}. After embedding, the logical local field $h_{i}$ and the couplings $J_{ij}$ are respectively distributed among the vertices and edges of the subgraph $S_{i}$ such that

\begin{equation*}
 \sum_{i_k \in V_{S_{i}}} h_{i_k} = h_{i}, \qquad  \sum_{i_k j_l \in E_{S_i}} J_{i_k j_l} = J_{ij}\,.
\end{equation*}
The parameter-setting problem consists of determining the qubit biases $h_{i_k}$ and coupler strengths $J_{i_k j_l}$ of the embedded problem.

The encoding process is of particular importance since the performance of the quantum processor is highly sensitive to the choice of embedding and its corresponding parameters. In what follows, we discuss both aspects of the encoding process, presenting the current approaches and describing the methods used in this work.

\subsubsection{Embedding Selection}	
\label{subsec:embedding_selection} 
The problem of finding an ME is in general NP-hard. In this work, we use the embedding software provided by D-Wave Systems. An ME found by this algorithm is not necessarily unique or optimal. In fact, different MEs representing the same QUBO problem might result in different hardware performance. Hence, a criterion or method to select an embedding which maximizes the hardware's performance is needed. We refer to this problem as the \textit{embedding selection problem}. 

Various embedding properties have been suggested as contributing factors affecting the hardware's performance. For instance, Cai et al. \cite{Cai14-ME} advise that it is preferable to select an embedding which minimizes either the total number of physical qubits used or the maximum number of physical qubits representing a logical qubit (i.e., it minimizes the length of the longest chain). Also, Pudenz et al. \cite{Pudenz2014} show that for the problem of antiferromagnetic chains, the performance of the hardware decreases with chain size. However, for some problems with a more complex topology, such as operational planning problems \cite{Rieffel15-Planning}, the chain length in the embedding did not contribute significantly to the hardware's performance. Further evidence that these embedding properties do not necessarily affect the performance of the device has been presented by Bian et al. \cite{Bian2016}. They introduce a locally structured embedding algorithm which considers the local structure of the hardware, in contrast to an ME approach which is referred to as a global embedding. In particular, they show that the local embedding, which requires more qubits and longer chains in general, yields better performance compared with the ME approach. Another embedding property has been considered in Ref. \cite{Boothby2016}, in which the authors suggest that it is preferable to use embeddings with chains of equal length. Considering these embedding properties as criteria in selecting an embedding has been sufficient for some specific problems. However, for general problems, a more sophisticated embedding selection strategy is still lacking.

Given that none of the current embedding selection criteria seem to predict or optimize the performance of the hardware for general problems, we implement an empirical approach to select an embedding. We refer to the embedding selected by this method as \textit{empirical embedding}. Our empirical selection method consists of generating $50$ distinct embeddings for each problem instance and setting their parameters according to the method described in Section~\ref{subsubsec:parameter_setting}. We run $1000$ anneals at $t_{a}=5$ $\mu \text{s}$ and use majority vote decoding as described in Section~\ref{sec:decoding} for each embedded instance. From this pool of embeddings, we select the $5$ embeddings, which results in a higher success probability. The experiment is repeated on the reduced pool of embeddings. We then select the embedding that maximizes the success probability. A similar approach has been used by King et al. \cite{King14-Algorithm}. To empirically select the embedding, we use a priori knowledge of the optimal value. In general, a different approach, which selects the solutions with minimum energy, could be considered instead. For example, Perdomo-Ortiz et al. \cite{Perdomo15-Pielite} introduce a performance estimator based on minimum-energy solutions to guide the selection of the best programming specifications. We use the optimal known solution to better demonstrate that the current embedding selection criteria do not necessarily optimize the performance of the hardware.

\subsubsection{Parameter Setting}
\label{subsubsec:parameter_setting}
A problem graph describing the structure of an optimization problem, such as the conflict graph $G_{\mbox{c}}$, has been reduced accurately 
to its minor-embedding $G_{\mbox{\tiny{emb}}}$ as long as there is a one-to-one correspondence between the ground states 
of the optimization problems defined on both graphs. It is well known that regardless of how the logical parameters, 
including local fields and couplers, are distributed among the physical qubits, a large 
negative value of the ferromagnetic couplers, $F$, ensures the correspondence between the ground states 
such that there is no broken connected subgraphs (i.e., the physical qubits encoding one 
logical qubit result in an identical state). However, since a quantum annealer is an analogue machine with finite available precision, it is 
important to find a sufficiently small value for these ferromagnetic couplers. 
In this case, the values of the ferromagnetic couplers are also dependent on how the logical parameters are 
distributed. An additional restriction of the DW2X architecture is the limited range of the $h$ and $J$ parameters, i.e., $[-2,2]$ and $[-1,1]$, respectively. The values of $h$ and $J$ can be scaled so that they lie within their respective ranges by multiplying them with a positive constant factor $\alpha$. This parameter is called the ``scaling factor'' in the rest of the paper. 

The literature on setting the parameters of an embedded graph can be divided, in general, 
into empirical and theoretical schemes. The former are the most widely used 
approaches, and include the work of Venturelli et al. \cite{Venturelli15-SKM}, Vinci et al. \cite{Vinci15-Error}, 
and Perdomo-Ortiz et al. \cite{Perdomo15-Pielite}. In the empirical approaches, the logical field and 
coupler values are evenly distributed among the physical qubits. To determine the strengths of 
the ferromagnetic couplers, the sample data from the device is used to experimentally find the 
optimal strength of the ferromagnetic couplers such that the probability of observing the 
ground state is maximized. The drawbacks of empirical approaches are twofold. First, inferring 
the strength of the ferromagnetic couplers based on the samples from 
a device that is subject to various types of errors does not guarantee that the ground state of the input graph $G_{\mbox{c}}$ 
is the same as the ground state of the embedded graph $G_{\mbox{\tiny{emb}}}$. Second, the empirical 
approaches are essentially trial-and-error approaches requiring significant computation time that 
increases as the size of the input graph increases. 

To the best of our knowledge, there is only one paper that discusses a theoretical approach to 
setting the parameters of an embedded graph \cite{Choi08-Parameter}. In the approach proposed 
by Choi \cite{Choi08-Parameter}, the distribution of the logical coupler values is similar to the empirical 
approaches, though the logical local field values are differently distributed. Furthermore, it is assumed 
that the subgraphs representing each logical qubit are subtrees of the hardware graph. However, this assumption does not necessarily 
hold for all embedded real-world problems, including some molecular similarity problems. 

In this work, we use a generalization of Choi's theoretical method and compare it with an empirical approach to 
investigate the potential benefit of a theoretical parameter-setting approach. It is worth mentioning that the empirical 
approach used in this paper is different from the approaches used in the literature. 
Both are explained below and the results of their performance are discussed 
in Section~\ref{subsection:emp_theo_parameter_setting}.

{\textbf{Empirical Parameter-Setting}} We empirically set the strengths of ferromagnetic couplers using the Hamze--de Freitas--Selby (HFS) algorithm \cite{Hamze04-HFS,Selby14-HFS}. Initially, we set $F_i = -1, ~\forall i \in V(G)$, where $F_i$ refers to the strength of all ferromagnetic couplers tying together the physical qubits representing the logical qubit $i$. We distribute the logical local field $h_{i}$ and couplings $J_{ij}$ evenly among the corresponding physical qubits and couplers, and scale their values to be in the available range. We then solve the scaled problem with the HFS algorithm $\mbox{iter}_{\tiny{\mbox{HFS}}}$ times. If there is no broken connected subgraph in at least one of the solutions, we stop and set the ferromagnetic coupler strengths to $-1$. Otherwise, we decrease $F_i$ by $\varepsilon > 0$ and repeat the above procedure. The empirical strength of ferromagnetic couplers used is the first value found for which none of the connected subgraphs are broken. We use $\varepsilon = 0.5$ and $\mbox{iter}_{\tiny{\mbox{HFS}}} = 5$ in our experiment.

{\textbf{Theoretical Parameter-Setting}} To theoretically set the $F_i$ values, we generalize the 
approach proposed by Choi~\cite{Choi08-Parameter}, where the logical couplings are distributed evenly among 
the physical couplers and the logical local fields are distributed following the procedure in Algorithm~\ref{parameter}. In our generalization, there is no need to reduce the subgraphs to 
trees. The details of our generalization are explained in Appendix \ref{subsec:appendix_theoretical_parameter_setting}.

\subsection{Decoding Strategies} 
\label{sec:decoding} 
A decoding strategy refers to the process of inferring the value of each logical qubit given 
the values of the physical qubits obtained from the quantum annealer. If the values of the physical qubits 
encoding one logical qubit agree, the common value of the physical qubits is assigned to the logical qubit. 
Otherwise, the logical qubits are considered broken and should be repaired. There are two general 
decoding strategies known as local and global \cite{Vinci15-Error}. The former approach decodes the broken logical qubits individually. Examples of local algorithms include heuristic algorithms with polynomial time complexity such as majority vote, coin tossing, and greedy descent. The latter approach decodes the broken qubits simultaneously. This process involves solving an optimization problem induced by the broken logical qubits, which is referred to as a ``decoding'' Hamiltonian. The disadvantage of a global decoding technique is that it requires solving an Ising problem belonging to the class of NP-complete problems. Examples of such algorithms are exhaustive search and simulated annealing.

Majority vote (MV), which assigns the most repeated value within an encoded subgraph to its 
corresponding logical qubit, is the simplest and most widely used approach in the literature 
\cite{Rieffel15-Planning,Perdomo15-Pielite,King14-Algorithm}. However, Vinci et al. \cite{Vinci15-Error} point out that 
MV is not appropriate for the minor-embedded problems since physical excitations are more 
likely to occur during the evolution of embedded problems and the probability that physical qubits break 
independently (the required criterion for 
the success of MV) cannot be considered valid. Furthermore, they mention that MV cannot be more effective than 
coin tossing in cases where the breakage of an encoded subgraph in various ways results in 
the same energy. Instead, they use simulated annealing to globally minimize the energy of the 
broken qubits. The main drawback of the global decoding strategies is that minimizing the 
energy of the decoding Hamiltonian can be as hard as solving the original Hamiltonian. To 
justify the computational cost, Vinci et al. \cite{Vinci15-Error} use the critical probability, $p_{\mbox{c}}$, 
as a decoding threshold. The critical probability is the probability of an infinite cluster appearing 
for the first time in an infinite lattice. If the probability for an encoded (logical) qubit to be broken, $p_{\tiny{\mbox{BQ}}}$, is greater than 
$p_{\mbox{c}}$, global approaches cannot be performed efficiently since the probability of having infinitely large broken clusters 
approaches 1 as the total number of logical qubits goes to infinity ($N \to \infty$). However, if $p_{\tiny{\mbox{BQ}}} < p_{\mbox{c}}$, 
the probability of having infinitely large broken clusters approaches 0 if the largest size of the decodable connected 
logical qubits is close to $\log(N)$.  Vinci et al. \cite{Vinci15-Error} show that the percolation threshold of their encoded 
logical graph is well beyond the experimental probability that a logical qubit breaks \cite{Vinci15-Error}. 
 
In this paper, we use MV to decode the broken hardware answer, map it back to the logical 
space, and then apply a greedy descent method to the logical answer as a post-processing technique to 
further refine the retrieved solution. Justification for the selection of MV as a local decoding strategy can be found in Section~\ref{subsec:greedy_descent}.

\section{Experimental Results}
\label{sec:experimental_results}

The quantum annealer used in this work is a DW2X processor located at NASA's Ames Research Center. It
consists of 1097 working flux qubits and 3060 working couplers arranged in a Chimera graph architecture and 
operates at 15 mK. To empirically evaluate the performance of a stochastic solver such as the DW2X, the standard 
metric is time-to-solution (TTS), the total time required by the solver to find the ground state at least 
once with a probability of $0.99$. Defining $\mbox{R}_{99}$ as the number of runs required by the solver to find 
the ground state at least once with a probability of $0.99$, we have $\mbox{TTS} = \mbox{R}_{99} \times \, t_a$, where $t_a$ 
is the duration of each annealing run. The calculation of TTS is explained in detail in 
Appendix \ref{subsec:appendix_TTS}. 

In this section, we first discuss the experimental setup. We then present the TTS results and investigate the effect of different 
embedding criteria, parameter setting approaches, and annealing times on the performance of the DW2X. 
Finally, we justify the use of MV as a decoding strategy and report on the improvement made on the performance of the
DW2X using a simple post-processing technique.

\subsection{Experimental Setup}
\label{subsec:experimental_setup}
We generate a parametrized family of the molecular similarity problems by using molecules from an available data set that consists of $7617$ small molecules \cite{Xu12}. The molecular similarity problem corresponds to finding the maximum weighted co-$k$-plex in a conflict graph $G_{\mbox{c}}$ as defined in Section \ref{sec:molecular_similarity}. The vertices $V$ and edges $E$ of $G_{\mbox{c}}$ represent the matchings and conflicts between the molecules being compared, respectively. The size of $G_{\mbox{c}}$, defined in terms of its number of vertices $|V|$, depends on the size of the molecules and the similarity conditions considered. It is worth mentioning that the number of logical variables in the molecular similarity problem corresponds to the size of $G_{\mbox{c}}$, not the size of the molecules being compared. The size of $G_{\mbox{c}}$ grows as the value of $k$ in Formulation \eqref{eq:co-k-plex_QUBO} increases. For example, the maximum size of $G_{\mbox{c}}$ that we can successfully embed into a DW2X processor is $|V|=46$ and $|V|=12$ for $k=1$ and $k=2$, respectively. In this paper, we consider $k=1$ in order to study the performance of the DW2X on larger non-native problems. The conflict graphs are then parametrized by $(V,d)$, where the number of vertices $|V|$ is selected from the set $\{ 18 + 4k | k=0, \ldots , 8 \} $, and the density $d$ is selected from the range in the set $\{ [ 65+ 10k , 75+ 10k ] | k=0, \ldots , 2 \} $. For each combination of $|V|$ and $d$, 100 different instances are generated.

\begin{figure}[h]
\begin{subfigure}{0.5\columnwidth}
    \centering
    \includegraphics[width=\columnwidth]{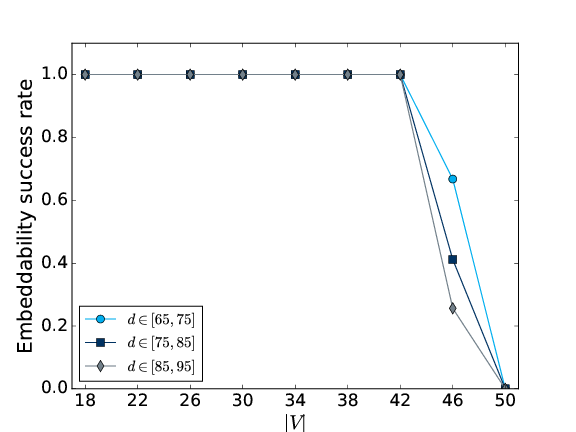}
    \caption{}
    \label{fig:phy_qubits_estimation:a} 
 \end{subfigure}
\begin{subfigure}{0.5\columnwidth}
    \centering
  \includegraphics[width=\columnwidth]{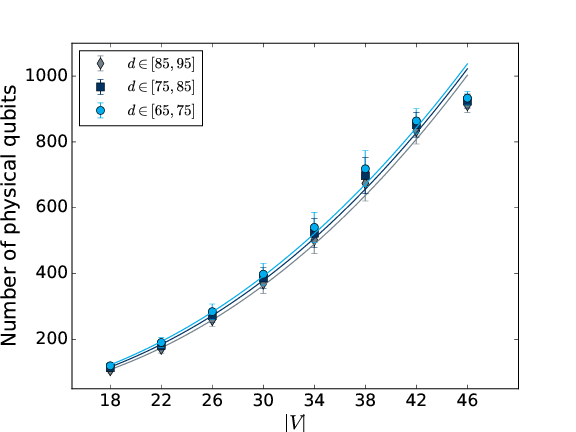}
  \caption{}
 \label{fig:phy_qubits_estimation:b}
 \end{subfigure}
\caption{Embeddability of graph similarity instances on the DW2X processor. 
(a) Fraction of the number of times an embedding was found. (b) Mean number of physical qubits 
used. The data points fit to $0.11x^{2.37}$, $0.14x^{2.32}$, and $0.17x^{2.27}$ for 
$d \in [65, 75]$, $d \in [75, 85]$, and $d \in [85, 95]$, respectively.}
\label{fig:phy_qubits_estimation}
\end{figure}

{\textbf{Embeddability}} To study the embeddability of the molecular similarity instances, we use the \texttt{find\_embedding()} function from SAPI 2.2 \cite{Cai14-ME} with default parameter values, but with the number of trials per function call set to $100$. Figure \ref{fig:phy_qubits_estimation:a} shows the mean success rate of the embeddability across 100 instances. The embeddability success rate is defined as the fraction of the number of embeddings found over the total number of trials. As shown in Figure \ref{fig:phy_qubits_estimation:a}, embeddings are found for all instances up to size $|V|= 42$, regardless of their densities. The number of embeddings generated decreases for size $|V|=46$ and eventually reaches 0 for size $|V|=50$. We also estimate the mean number of physical qubits required to embed the given instances. As illustrated in Figure \ref{fig:phy_qubits_estimation:b}, the mean number of physical qubits scales similarly across different densities. The scaling is $\text{O}(|V|^{2.37})$, $\text{O}(|V|^{2.32})$, and $\text{O}(|V|^{2.27})$ for density $d \in [65, 75]$, $d \in [75, 85]$, and $d \in [85, 95]$, respectively.

\subsection{Time-to-Solution}
\label{subsection:TTS} 
\begin{figure}[h!]
  \centering
  \subcaptionbox{$d \in [65, 75]$}[.4\linewidth][c]{%
    \includegraphics[width=.4\linewidth]{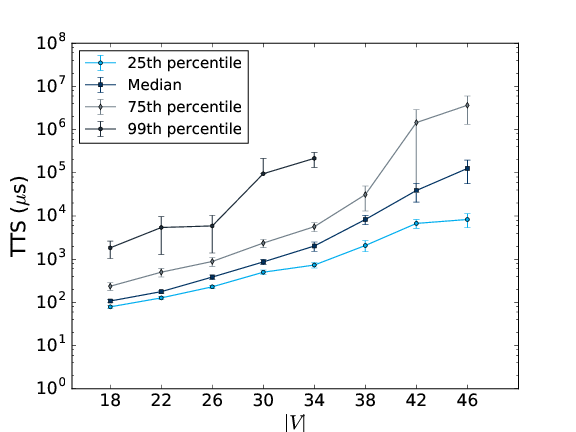}}\quad
  \subcaptionbox{$d \in [75, 85]$}[.4\linewidth][c]{%
    \includegraphics[width=.4\linewidth]{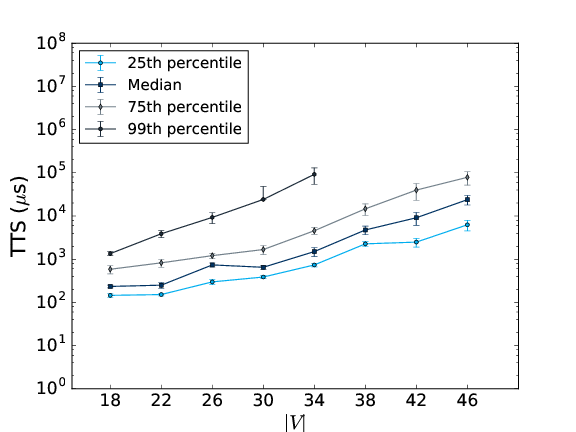}}\quad
  \subcaptionbox{$d \in [85, 95]$}[.4\linewidth][c]{%
    \includegraphics[width=.4\linewidth]{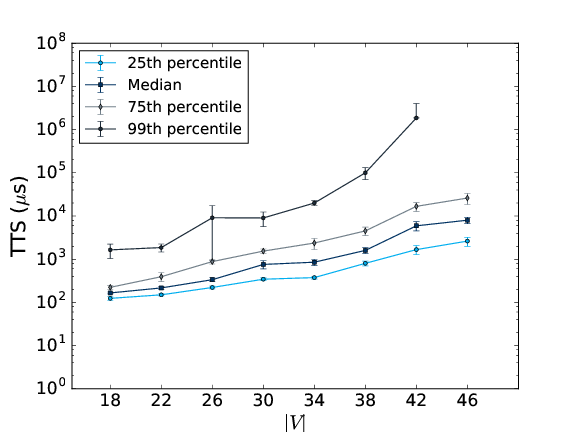}}
\caption{Different percentiles of the TTS (logarithmic scale) as a function of the
number of logical variables for (a) $d \in [65, 75]$, (b) $d \in [75, 85]$, and (c)
$d \in [85, 95]$. The number of calls to the DW2X and the number of anneals per call are 5 and 10,000, respectively.}
\label{fig:TTS_main}
\end{figure}

Figure \ref{fig:TTS_main} shows the 25th, 50th, 75th, and 99th percentiles of the TTS for three different densities over 
100 instances for each problem size. For every instance, the true ground state is found using the Gurobi solver (version 6.5.1), 
the best empirical embedding is selected, the parameters are theoretically set, the annealing time is set at $20$ $\mu s$ 
(see Section~\ref{subsec:annealing_time}), the retrieved solution from the hardware is decoded using MV. A common strategy to average out systematic errors is to perform gauge transformations between each call to the quantum annealer \cite{Boixo14}. Since we are plotting the TTS for a class of instances of similar size, not an individual instance, and the TTS has a higher variance over different instances than different gauges, the timing data is not averaged over multiple gauges \cite{Hen15-Probing}.

The 99th percentile of the TTS is not shown in Figure \ref{fig:TTS_main} if all 100 instances are not solved given 
5 calls to the DW2X and 10,000 anneals per call. The lower limit of the standard deviation of the 99th percentile is also not shown for 
several sizes in case it is bigger than the mean of the 99th percentile. As illustrated, the TTS is higher for 
problems with lower densities. As the density of the QUBO graph reaches the extreme values 
of 0 and 100, we intuitively expect that the problem of identifying the maximum independent 
set becomes easier. Furthermore, the large differences between the median and 99th percentiles of the TTS for 
several sizes and densities are indications of heavy tails in the distributions of the TTS. In other words, this result 
conveys that the hardness of our problems does not depend solely on the number of logical variables. 
It significantly differs across instances with a similar number of logical variables.

\subsection{Comparing Embedding Selection Criteria}

\begin{figure}[h!]
\centering
\includegraphics[scale=0.4]{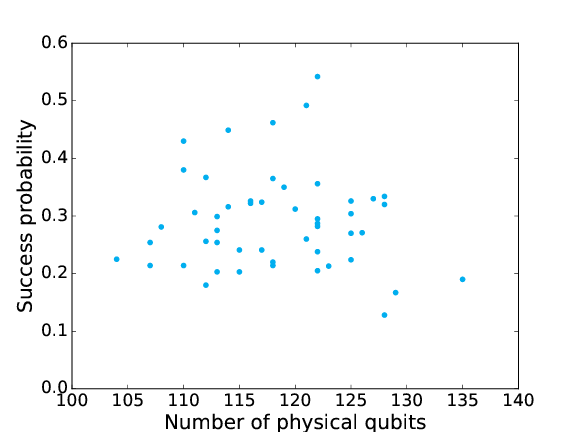}
\caption{Success probabilities for an instance of the molecular similarity problem of size $|V| = 18$ and density $d \in [75, 85]$. The results are shown for $50$ different embedding representations of the same instance.}
\label{fig:emb_motivation}
\end{figure}

Following the discussion in Section~\ref{subsec:embedding_selection}, different embedding representations of the same Ising Hamiltonian might result in different performance of the QA hardware. In the literature, it has been suggested that optimizing some embedding properties could be an important factor that affects the performance of the hardware, for example, minimizing the total number of qubits used in an embedding. Figure \ref{fig:emb_motivation} depicts a counterexample to this argument for an instance of the molecular similarity problem. In particular, we do not observe any type of correlation between the mean success probability and the number of physical qubits used in the embedding. Additionally, we note that if the embedding yielding a low success probability is selected, such an instance will be misclassified as a hard instance and might mask a heavy tail in the final TTS results. These observations motivate us to perform a more comprehensive study to address the embedding selection problem. 

To test the overall impact of the embedding selection problem on the quantum hardware's performance, we experimentally compare three criteria previously proposed in the literature. We generated $100$ instances for each problem size and $50$ embeddings for each instance, and selected $1$ embedding according to each of the criteria. Specifically, we consider the following criteria:
\begin{enumerate}
\item Selecting the embedding with the smallest number of physical qubits (PQ);
\item Selecting the embedding with the shortest longest chain (LCh);
\item Selecting the embedding with chains of equal length (STD).
\end{enumerate}
Whereas the first two criteria are straightforward, heuristic ME algorithms do not necessarily return embeddings with chains of equal length as required for the third criterion. To account for the selection of an embedding with chains of equal size, we calculate the standard deviation of the chain size for each embedding and select the embedding with the minimum standard deviation. Further, we consider the empirical selection criteria for which we select the embedding that yield the highest success probability.
\begin{figure}[ht] 
  \begin{subfigure}[b]{0.5\linewidth}
    \centering
    \includegraphics[width=0.75\linewidth]{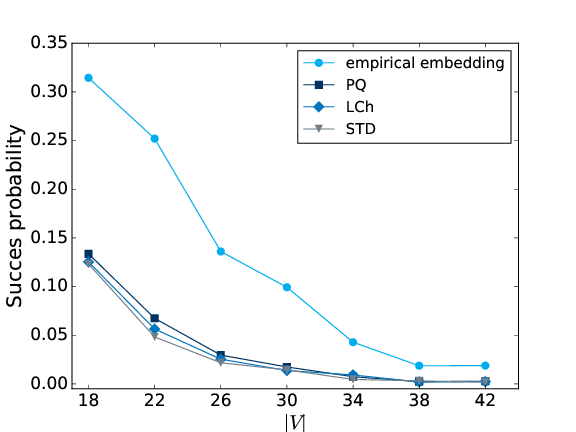} 
    \caption{Embedding selection comparison} 
    \label{fig:emb_selec:a} 
    \vspace{4ex}
  \end{subfigure}
  \begin{subfigure}[b]{0.5\linewidth}
    \centering
    \includegraphics[width=0.75\linewidth]{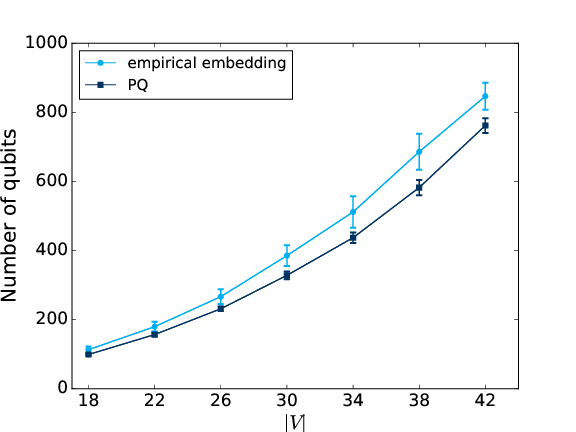} 
    \caption{Mean number of qubits used} 
    \label{fig:emb_selec:b} 
    \vspace{4ex}
  \end{subfigure} 
  \begin{subfigure}[b]{0.5\linewidth}
    \centering
    \includegraphics[width=0.75\linewidth]{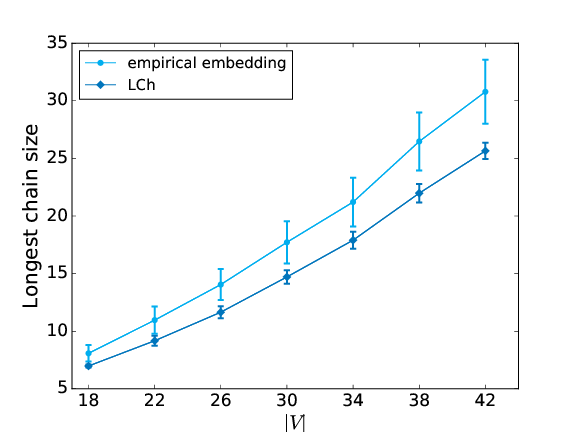} 
    \caption{Mean size of longest chain} 
    \label{fig:emb_selec:c} 
  \end{subfigure}
  \begin{subfigure}[b]{0.5\linewidth}
    \centering
    \includegraphics[width=0.75\linewidth]{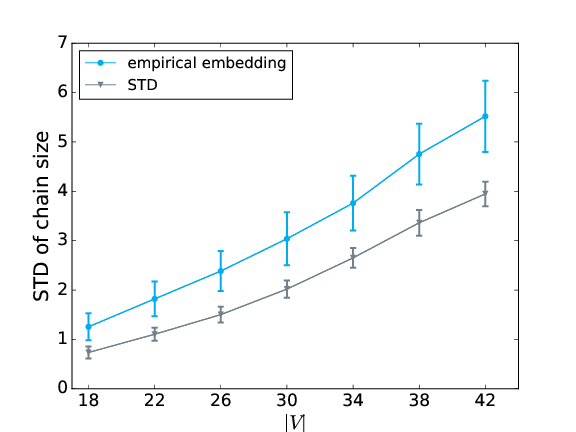} 
    \caption{Mean STD of chain size} 
    \label{fig:emb_selec:d} 
  \end{subfigure} 
  \caption{Comparison of different embedding selection criteria for instances with logical density $d \in [75, 85]$. Each point indicates the mean across $100$ instances of each problem size. a) shows the comparison in terms of the mean success probability for the DW2X device performing $1000$ annealing runs. Additional results show comparison in terms of (b) the mean number of physical qubits, (c) the mean size of the largest chain, and (d) the mean STD of chain size.}
  \label{fig:emb_selec} 
\end{figure}
Figure~\ref{fig:emb_selec:a} shows the embedding selection comparison for the three criteria and the empirical embedding in terms of the mean success probability. We observe that these three selection criteria perform similarly poorly against the best embedding found empirically. Further, the empirical embedding uses on average more physical qubits, its longest chain is longer, and its chains have higher STD as shown in Figures \ref{fig:emb_selec:b}, \ref{fig:emb_selec:c}, and \ref{fig:emb_selec:d}, respectively. Therefore, the use of these criteria has no effect on the success probability.

An additional observation can be made from Figure~\ref{fig:emb_selec:a}. As the logical problem size increases, the difference between the embedding selection methods decreases. A possible reason could be that for larger problems, most of the embeddings found with the heuristic algorithm reach the limit size of the chip. Thus, we have less freedom to generate different embeddings. This observation is in agreement with the argument provided by Cai et al. \cite{Cai14-ME} for the success of a heuristic ME algorithm. Specifically, they state that if the problem to be embedded is significantly smaller than the hardware graph's size, there are probably more--distinct embeddings. 

To generate the results shown in Figures \ref{fig:emb_motivation} and \ref{fig:emb_selec}, we perform $1000$ annealing runs for each problem instance. In order to verify if performing gauge averaging or a larger number of anneals will affect our conclusions regarding the embedding selection problem, we perform additional experiments. Specifically, we repeat the test shown in Figure~\ref{fig:emb_motivation} for two cases: $1000$ anneals and $5$ random gauges; and 50,000 anneals and the default gauge. We also repeat the test for the first criterion presented in Figure~\ref{fig:emb_selec:a} using $1000$ anneals and 5 random gauges. In all cases, the new results verify that there is no correlation between the number of physical qubits used and the success probability.

These results suggest that further investigation is needed in order to understand which properties of embeddings have an influence on the hardware's performance when solving problems with a complex topology, such as the conflict graphs in the molecular similarity problems. Meanwhile, an empirical method like the one performed here can be applied.  

\subsection{Empirical versus Theoretical Parameter Setting}
\label{subsection:emp_theo_parameter_setting}

\begin{figure}[ht] 
  \begin{subfigure}[b]{0.5\linewidth}
    \centering
    \includegraphics[width=0.75\linewidth]{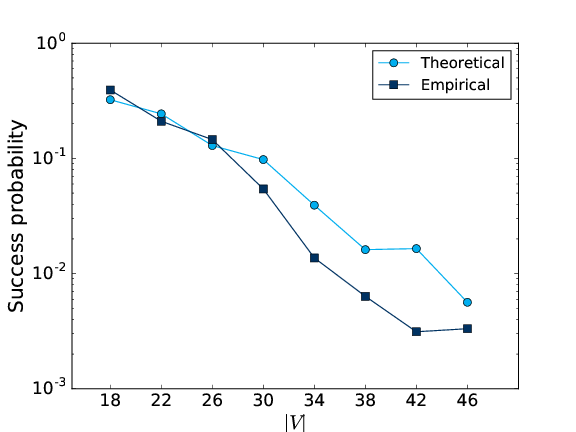} 
    \caption{Success probability}
    \vspace{4ex}
  \end{subfigure}
  \begin{subfigure}[b]{0.5\linewidth}
    \centering
    \includegraphics[width=0.75\linewidth]{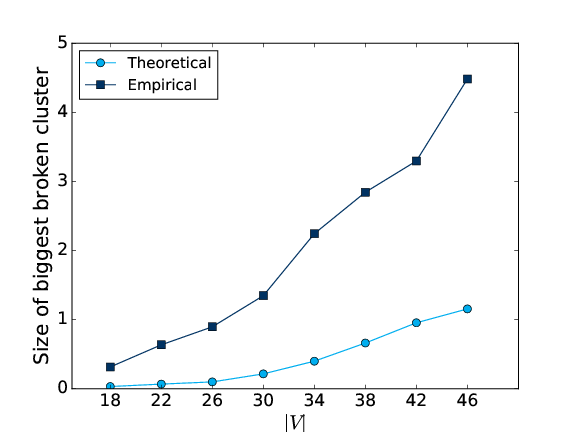} 
    \caption{Size of biggest broken cluster}
    \vspace{4ex}
  \end{subfigure} 
  \begin{subfigure}[b]{0.5\linewidth}
    \centering
    \includegraphics[width=0.75\linewidth]{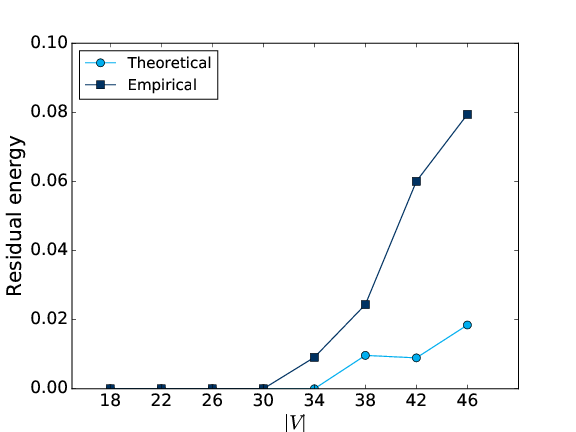} 
    \caption{Residual energy} 
  \end{subfigure}
  \begin{subfigure}[b]{0.5\linewidth}
    \centering
    \includegraphics[width=0.75\linewidth]{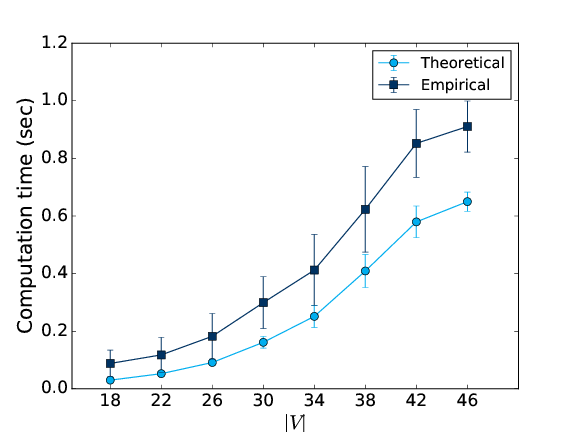} 
    \caption{Computation time} 
  \end{subfigure} 
  \caption{Parameter setting comparison for molecular similarity instances with $d \in [75, 85]$. 
(a) shows the success probability (log) when the parameters of the embedded graphs are set theoretically 
or empirically. (b) shows the size of the biggest broken cluster size for theoretical and empirical 
parameter settings. (c) illustrates the residual energy for theoretical and empirical parameter settings. (d) 
illustrates the computation time (seconds) to set the parameters theoretically and empirically.
All four measures represent the average value over 100 instances. These figures are generated by setting 
$t_a=5$ $\mu s$, and the number of calls and anneals per call are set to 5 and 10,000, respectively.}
 \label{fig:theor_empirical} 
\end{figure}

Figure \ref{fig:theor_empirical} shows the comparison of the performance of the DW2X using two 
different parameter-setting schemes with respect to four measures: success probability, size of 
the biggest broken cluster, residual energy, and computation time. The second measure represents the largest size of 
a decodable domain in which all broken logical qubits are connected. The third measure is the relative 
energy difference between the quantum annealer's lowest-energy solution and the actual ground state 
found by the Gurobi solver.

As illustrated, the theoretical approach is overall superior to the empirical approach. It 
results in a higher probability of success (especially as the size of the problem increases), smaller 
broken clusters, lower residual energy, and lower computation time. These results indicate that the theoretical approach 
can be a strong alternative to replace current empirical practices in the literature. 
The theoretical approach reduces the pre-processing computation time to set the problem parameters and boosts the performance of the quantum annealer.

It is worth mentioning that the reported performance for 
the empirical approach in this paper is most likely an upper bound on the performance of the 
typical empirical approaches in the literature for two main reasons. First, the value of the 
ferromagnetic couplers is optimized per instance in our empirical approach, whereas it is usually 
optimized per class of instances with the same size in the literature. Second, in our approach, the 
analogue, noisy quantum solver is replaced by a classical solver with higher precision, thereby 
decreasing the probability that the qubits representing one logical qubit resolve into different states.

\begin{figure}[h!]
\centering
\includegraphics[scale=0.4]{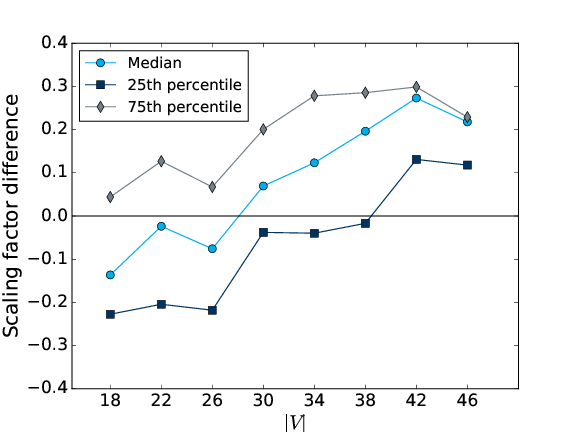}
\caption{Comparison of the 25th, 50th, and 75th percentiles of the difference between the theoretical scaling factor and the empirical one. 
The number of calls to the quantum annealer is 5, the 
number of anneals per call is set to 10,000, and $t_a=5$ $\mu s$.}
\label{fig:theor_empirical_scalefactor}
\end{figure}

To gain more insight into the nature of the superiority of the theoretical parameter-setting approach, the difference between 
the theoretical scaling factor and the empirical scaling factor for each instance was found. Figure \ref{fig:theor_empirical_scalefactor} shows the 
$25$th, $50$th, and $75$th percentiles of the scaling difference over 100 instances for each problem size. As illustrated, the theoretical approach yields a larger scaling factor 
as the number of logical variables increases (the difference is positive), which explains its better performance over the empirical approach.

\subsection{Annealing Time}
\label{subsec:annealing_time}

Solving classical optimization problems by exploiting quantum effects is of central interest in the field of experimental quantum annealing. Although the presence of entanglement \cite{Lanting14-Entanglement} and multiqubit tunnelling \cite{Boixo16-Tunneling} in D-Wave processors have been experimentally demonstrated, any scaling advantage over classical algorithms remains elusive. When studying scaling, it is important to determine an optimal implementation. In the case of non-native problems, optimality refers to determining an effective encoding, as has been addressed in the previous sections. Once an encoding has been determined, it is necessary to find the optimal annealing time. 

In Ref. \cite{Ronnow2014}, it was suggested that the annealing time must be optimized for each problem size. In this context, various studies have tried to determine an optimal annealing time for different random instances and have arrived at the conclusion that the minimum possible $t_{a} = 20$ $\mu \text{s}$ of D-Wave processors is longer than the optimal annealing time \cite{Denchev15-Google,Ronnow2014}. Alternatively, Hen et al. \cite{Hen15-Probing} have found instances whose optimal annealing time on the same processor is greater than $ t_{a} = 20$ $\mu \text{s}$. Although it is outside of the scope of this paper to probe for quantum speedup, we are interested in studying the DW2X's performance dependance on the annealing time for our molecular similarity problem instances. A recent upgrade to the DW2X processor introduced a faster minimum annealing time of $t_{a} = 5$ $\mu \text{s}$, allowing us to test the quantum annealer's performance dependency on $t_{a}$ for shorter annealing times. 

\begin{figure}[h!]
\centering
\includegraphics[scale=0.4]{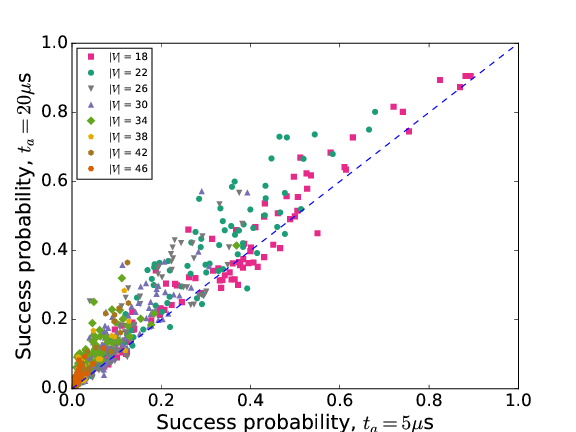}
\caption{Success probability correlation for the DW2X at $t_{a} = 5$ $\mu \text{s}$ and $t_{a} = 20$ $\mu \text{s}$. Results are shown for different numbers of logical variables, varying from $|V| = 18$ to $|V| = 46$ and coloured accordingly, and density $ d \in [75, 85]$.}
\label{fig:sp_correlation}
\end{figure}

From the correlation plot in Figure \ref{fig:sp_correlation}, it seems that a longer annealing time increases the success probability. In order to gain better insights, we performed a \textit{t}-test to determine whether the difference between these samples is statistically significant. We obtained a $p$-value close to $0$, which implies that the higher mean success probability at $t_{a}= 20$ $\mu \text{s}$ has not occurred only by chance. This result provided enough evidence to accept the hypothesis that the difference between the mean success probabilities at $t_{a}=20$ $\mu \text{s}$ and $t_{a}=5$ $\mu \text{s}$ is statistically greater than $0$, which can be understood in two different ways. Firstly, it has been suggested that the success probability is a monotonically increasing function of $t_{a}$ \cite{Ronnow2014}. In this sense, we should expect that longer annealing times will increase the success probability until the asymptotic regime is reached. A counterexample to this assumption has been presented by Amin \cite{Amin2015}, who presented a nonmonotonic success probability function of $t_{a}$. According to his findings, our results for the success probability at $t_{a} = 5$ $\mu \text{s}$ and $t_{a} = 20$ $\mu \text{s}$ indicate that our system is in a quasistatic regime and that those annealing times are still too long. 

\subsection{Majority Vote as a Decoding Strategy}
\label{subsec:greedy_descent}

\begin{figure}[h!]
\centering
\includegraphics[scale=0.4]{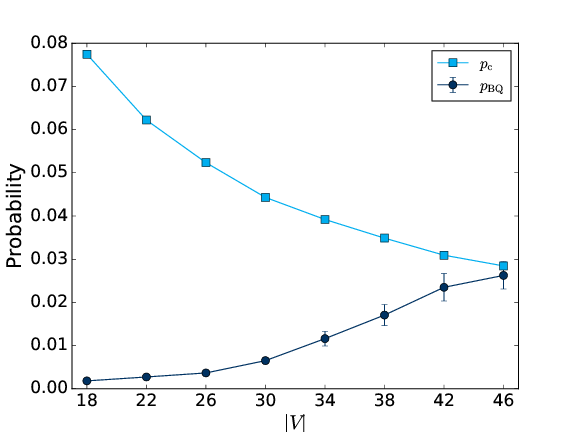}
\caption{Mean probabilities of broken qubits and mean percolation thresholds
for instances with different number of logical variables and $d \in [75, 85]$. The average is taken over 100 
instances.}
\label{fig:percolation_prob_all_sizes}
\end{figure}

Figure \ref{fig:percolation_prob_all_sizes} shows the mean experimental probability of there being broken 
qubits ($p_{_{\tiny{\mbox{BQ}}}}$) for different sizes of the molecular similarity instances with $d \in [75,85]$, as well as 
their corresponding mean percolation thresholds ($p_{\mbox{c}}$). The calculation of the percolation threshold is explained in 
detail in Appendix \ref{subsec:appendix_percolation}. As illustrated, the values of $p_{_{\tiny{\mbox{BQ}}}}$ approach $p_{\mbox{c}}$ as the number of logical variables increases. The trend of the $p_{_{\tiny{\mbox{BQ}}}}$ and 
$p_{\mbox{c}}$ values shows the experimental probability of there 
being broken qubits would eventually be larger than the percolation threshold for larger graph similarity 
instances. Therefore, the global decoding schemes will be inefficient for large instances of our graph 
similarity problem. The probability of having large broken clusters would be high, and solving the 
remaining problem would be as hard as solving the original.
  
\begin{figure}[h!]
\centering
\includegraphics[scale=0.4]{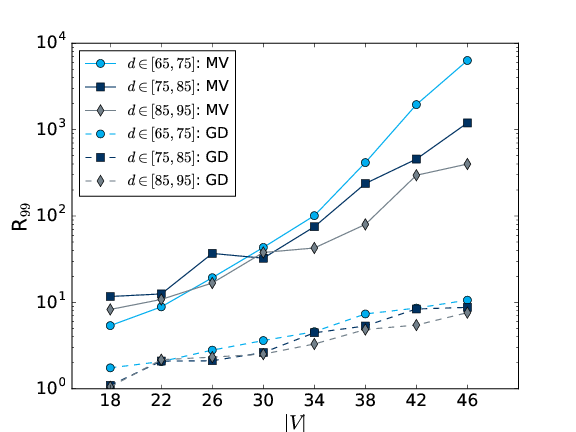}
\caption{Median $\mbox{R}_{99}$ for three different densities using 
MV and greedy descent, where the number of calls and anneals per call are set to 5 and 10,000, respectively.}
\label{fig:r99_MV_GD}
\end{figure}

To locally decode the physical solutions, we apply MV since it requires the least possible effort to retrieve a 
solution from the physical space. It also gives a clear picture of the efficiency of the quantum annealer, with minimal 
contribution from a classical computer. The decoded logical solutions obtained by MV might be local optima. To further improve the quality of 
the decoded solutions, we use the greedy descent method. Figure \ref{fig:r99_MV_GD} shows the effect of applying greedy descent on the 
decoded logical solutions obtained by MV. As shown, there is a noticeable difference between the median $\mbox{R}_{99}$ 
before and after applying greedy descent, implying that refining the quantum solutions via classical 
post-processing techniques would be necessary to efficiently solve non-native problems using a quantum annealer.

\section{Conclusions and Discussion} 
\label{sec:conclusion}

In this work, we have studied the performance of a quantum annealer in solving instances of the molecular similarity problem utilizing a real data set. The effective use of a quantum annealer presents many challenges. Here, we focused on challenges derived from encoding and decoding real-world problems. We addressed the challenges present in both aspects of encoding, that is, embedding and parameter setting, and demonstrated how a careful encoding strategy helps to improve the performance of the DW2X device. 

In particular, our results emphasized the importance of the embedding selection problem. For some instances, we observed that two different embedding representations of the same original problem yield very different success probabilities. Commonly, this difference is attributed to various properties of embedding, for example, the number of qubits used. However, we have shown that none of the commonly observed embedding properties correlate with the hardware's performance. In this work, we have incorporated an empirical method that selects the embedding that maximizes the success probability. We have also observed that the performance of the quantum annealer is less sensitive to the choice of embedding when increasing the size of the original problem. This result is expected since a minor-embedding heuristic will successfully find a larger number of distinct embeddings if the size of the problem to be embedded is significantly smaller than the size of the chip. Thus, we expect that the poor performance of embedded problems with a size as large as the size of the current chip should improve in a next-generation quantum annealer.

We have also shown that using a theoretical, rather than empirical, approach to select the parameters in the embedding can have significant advantages. Besides eliminating the impractical experimental penalty optimization, a theoretical parameter-setting approach ensures an accurate representation of the logical Ising Hamiltonian, and as a consequence reduces the probability of broken qubits. Of particular importance is the overall boosting of the hardware's performance, which benefits from an improved scaling factor over the empirical approach.  

Another important question we have addressed is the selection of a decoding technique. For our problem set, we have found that a simple local decoding technique is effective. This is mainly a consequence of the reduced connectivity of broken logical qubits. Specifically, we have shown that fufrther improvement in the hardware performance can be achieved if we use majority vote to fix the broken solutions and subsequently apply a greedy descent post-processing technique.

Whereas our experimental conclusions are restricted to the specific type of problem studied here, the results provide useful insight for future non-native benchmarking studies.

In future work, we will address some of the questions still open regarding the effective implementation of a problem embedded into a quantum annealer. Of practical importance is gaining a better understanding of which properties correlate to the performance of the quantum annealer. One possible research direction could be to study the quality of the physical qubits and couplers in the chip.\footnote{H. Katzgraber, personal communication, 2016.} An alternative direction could be to study whether the properties defining the hardness of spin glass problems have an impact on the hardware's performance. After having determined an optimal encoding, we will address the problem of determining the optimal annealing time.




\appendix
\section{Appendix}

\subsection{Theoretical Parameter Setting}
\label{subsec:appendix_theoretical_parameter_setting}

As mentioned in Section \ref{subsubsec:parameter_setting}, we generalize the approach presented by 
Choi \cite{Choi08-Parameter} to theoretically set the parameters of the embedded graph, in which the connected subgraphs representing the logical qubits are not necessarily subtrees of 
the hardware graph. The procedure used in this paper is explained in Algorithm \ref{parameter}. 
It is worth mentioning that we do not discuss the proof for the validity of the theoretical 
approach. The key idea as presented by Choi \cite{Choi08-Parameter} is to ensure that the ground state 
of the input graph before and after embedding matches. 

Let us consider a conflict graph $G_{\mbox{c}} =(V_{\mbox{c}}, E_{\mbox{c}})$ with logical local fields and couplings denoted by 
$h$ and $J$, respectively. Assume that logical qubit $i \in V_{\mbox{c}}$ is represented by $n_i$ physical 
qubits forming the physical subgraph $S_i = (V_{S_i}, E_{S_i})$. Further assume that 
$\mbox{lnb}(i)$ is the set of neighbouring vertices of the logical qubit $i$ in the logical graph $G_{\mbox{c}}$ and 
$\mbox{pnb}(l)$ is the set of neighbouring vertices of the physical qubit $l \in V(S_i)$,  $\forall i \in V_{\mbox{c}}$, 
excluding the vertices representing the same logical qubit. 

The theoretical approach detailed below has an iterative pre-processing step in which several 
logical qubits might be removed from the logical graph, since their optimal values can be 
inferred in advance. The parameters are then set on the reduced logical graph. 

\begin{algorithm}[h!]
\caption{Set the parameters of the embedded graph}
\label{parameter}
\begin{algorithmic}[1]
\For {logical qubit $i \in V_{\mbox{c}}$}
	\State {calculate a new parameter $C_i = \sum_{j \in \text{lnb(i)}} |J_{ij}| - |h_i|$}
\EndFor 
\While {there is at least one $C_i < 0$}
	\State {remove any logical qubit $i$ with $C_i <0$ and set its value 
			to $+1$ if $h_i <0$, and at $-1$ otherwise}
	\State {denote the remaining graph by $G_{\mbox{c}} = (V_{\mbox{c}}, E_{\mbox{c}})$}
	\State {update the local fields of the remaining logical qubits}
	\State {calculate a new $C_i$ value for the remaining logical qubits}
\EndWhile
\State {distribute logical $J$ evenly among the physical couplings connecting two logical qubits}
\For {logical qubit $i \in V_{\mbox{c}}$} 
	\For {physical qubit $l \in V_{S_i}$}
		\State {set $h_l = sign(h_i) \big(\sum_{k \in \text{pnb(l)}} |J_{lk}| - \frac{C_i}{n_i} \big)$} 
	\EndFor
	\For {ferromagnetic physical coupler $e \in E_{S_i}$}
		\State {set $F_e =  - \frac{(n_i -1)}{n_i} C_i - \epsilon$ (we use $\epsilon = 0.1$)}
	\EndFor
\EndFor
\State find the scaling factor to bring the distributed $(h,J,F)$ to the range specified by the user (we use the range $(-0.8,0.8)$  
		for both $h$ and $J$ in this paper)
\State multiply the distributed $(h,J)$ by the scaling factor
\State set all ferromagnetic coupler values ($F$) to $-1$
\end{algorithmic}
\end{algorithm}

\subsection{Time-to-Solution Estimation}
\label{subsec:appendix_TTS}

Since the quantum annealer is a stochastic solver, we consider the successive annealing 
runs as a sequence of binary experiments that might succeed in returning the ground state 
with some probability. Let us formally define $X_1, X_2, \ldots, X_n$ as a sequence of random 
independent outcomes of $n$ annealing runs, where $\mathbbm{P}(X_i=1)=\theta$ denotes the probability of observing the ground 
state at the $i$-th anneal. Defining $Y$ as the number of successes observed in $n$ anneals 
($Y=\sum_{i=1}^{n} X_i$), we have $\mathbbm{P}(Y = y|\theta) = {n \choose y} (1-\theta)^{n-y}\theta^y$ ($Y|\theta \sim \mbox{Bin}(n, \theta)$). 
That is, $Y|\theta$ has a binomial distribution with parameters $n$ and $\theta$. The $\mbox{R}_{99}$ then equals $n$ such that 
$\mathbbm{P}(Y \geq 1|\theta) = 0.99$. It is easy to verify that $\mbox{R}_{99} = \log (1-0.99) / \log(1-\theta)$. Since 
the probability of success $\theta$ is unknown, the challenge is to estimate $\theta$.

We follow the Bayesian inference technique to estimate the probability of success for 
each instance $i$ \cite{Hen15-Probing}. In the Bayesian inference framework, we start 
with a guess on the distribution of $\theta$ known as prior and update it based on 
the observations from the device in order to get the posterior. Since the observations from the device 
have a binomial distribution, the proper choice of prior is a beta distribution which is the conjugate prior of the 
binomial distribution. This choice guarantees that the posterior also has a beta distribution. The beta distribution with parameters 
$\alpha=0.5$ and $\beta=0.5$ (the Jeffery prior) is chosen as prior since it is invariant under reparameterization 
of the space and it learns the most from the data \cite{Clarke94}. 

Updating the Jeffery prior based on the data from the device, the posterior distribution denoted 
by $\pi_i(\theta)$ is then
\begin{align}
\pi_i(\theta) \sim \mbox{Beta} \bigg(0.5+\sum_{c=1}^{C} y_{ci}, 0.5+NC-\sum_{c=1}^{C} y_{ci} \bigg), & \notag 
\end{align}
\noindent where $C$ is the number of calls to the quantum annealer, $N$ 
is the number of anneals in each call, and $y_{ci}$ is the number of times that the ground state 
of instance $i$ is observed at the $c$-th call. 

To estimate the TTS (or ${\mbox{R}}_{99}$) for the entire population of instances with 
similar parameters, let us assume that there are $I$ instances with similar 
properties, for example, with the same number of variables. We are interested in using the data on these 
$I$ instances to estimate the TTS for the entire population of instances 
with the same number of variables. After finding the posterior distribution $\pi_i(\theta)$ for all instances 
in set $\{I\}$, we use the bootstrap methodology to estimate the distribution of the $q$-th percentile of the TTS. 
The procedure is described in Algorithm~\ref{TTS}.

\begin{algorithm}[h!]
\caption{Estimate the distribution of the $q$-th percentile of the TTS}
\label{TTS}
\begin{algorithmic}[1]
\State fix the number of bootstrap re-samples $B$ (we set it at $1000$)
\For  {$b=1, \ldots, B$}
	\State {sample a new set of instances from the set $\{I\}$ with replacement and 
	    length $I$}
	  \For {each sampled instance $j$} 
	  	\State {sample a value from its corresponding 
	    			posterior probability distribution $\pi_{j}(\theta)$ to obtain the set $\{p_{jb}\}$}
	\EndFor
\EndFor
\For {$b=1, \ldots, B$} 
	\State{find the $(100-q)$-th percentile of set $\{p_{jb}\}$ and denote it by $p_{(1-q)b}$}
\EndFor
\For {$b=1, \ldots, B$}
	\State {estimate the $q$-th percentile of the TTS as 
	   $\mbox{TTS}_{qb} = \tau \log (1-0.99) / \log(1-p_{(100-q)b})$}
\EndFor
\State consider the empirical distribution of $(\mbox{TTS}_{q1}, \mbox{TTS}_{q2}, \ldots, \mbox{TTS}_{qB})$ as
		an approximation of the true $\mbox{TTS}_{q}$ distribution (we have plotted the mean and the standard 
		deviation of this empirical distribution in Figure \ref{fig:TTS_main})
\end{algorithmic}
\end{algorithm}

\subsection{Percolation Threshold of the Molecular Similarity Problem Instances}
\label{subsec:appendix_percolation}

The QUBO graphs of molecular similarity problem instances can be considered random graphs. A random 
graph is a collection of vertices with edges connecting pairs of them randomly. 
Newman et al. \cite{Newman01} and Callaway et al. \cite{Callaway00} developed an approach 
based on generating functions to determine the statistical properties of random 
graphs with arbitrary degree distribution. Here we review their approach to determine 
the percolation threshold. 

Let us denote the probability that a randomly chosen vertex has degree $k$ by $p_k$. Let us further define
\begin{align}
G_0(x) = \sum_{k=0}^{\infty} p_k x^k, ~~~~G_1(x) = \frac{1}{z} G^{\prime}_0(x),~~~~\mbox{and} 
~~~~z = G^{\prime}_0(1), & \notag
\end{align}
\noindent where $G_0(x)$ and $G_1(x)$ are the generating functions for the probability distribution 
of vertex degrees and outgoing edges, respectively, and $z$ is the average vertex degree. 
Callaway et al. \cite{Callaway00} have shown that the percolation threshold, or critical probability, can be calculated as follows:
\begin{align}
p_{\mbox{c}} = \frac{1}{G^{\prime}_1(1)}  = \frac{G^{\prime}_0(1)}{G^{\prime \prime}_0(1)}. & \notag
\end{align}
The details of the derivation can be found in \cite{Newman01,Callaway00}. The key idea is 
that the critical probability is the point at which the mean cluster size goes to infinity.

The context in which we apply this criterion is on random graphs whose degree distribution is known. 
It is known because the degree distribution can be measured directly \cite{Newman01}. Below, we provide an 
example with a known degree distribution for which we calculate the critical probability. 

\noindent \textbf{Example~} Consider a graph similarity problem instance with 18 vertices and $d \in [65, 75]$. The 
number of vertices with degree 11, 12, and 13 are, in respective terms, 8, 8, and 2.
The distribution of vertex degrees can be generated by 
\begin{align}
G_0(x) = \frac{8x^{11} + 8x^{12} + 2x^{13}}{18}. & \notag
\end{align}
Applying the formula above, the critical probability is $p_{\mbox{c}} = 0.0934$. 

\end{document}